\documentclass[12pt,a4paper,aps,rmp,onecolumn]{revtex4}
\usepackage{latexsym}
\usepackage{times}
\usepackage{graphicx}
\usepackage[tight,FIGTOPCAP]{subfigure}
\usepackage{amsmath,amssymb,enumerate,mathrsfs,boxedminipage,fancybox}

\newcommand{\eulang}[2]{\genfrac{\langle}{\rangle}{0pt}{}{#1}{#2}}
\newcommand{\su}[2]{\genfrac{(}{)}{0pt}{}{#1}{#2}}

\def\mezzo{\frac{1}{2}}

\def\s{\sigma}

\def\t{\tau}

\def\d{\delta}

\def\bi{\begin{itemize}}
\def\ei{\end{itemize}}

\def\be{\begin{enumerate}}
\def\ee{\end{enumerate}}

\def\beq{\begin{equation}}
\def\eeq{\end{equation}}

\def\bdm{\begin{displaymath}}
\def\edm{\end{displaymath}}

\def\bsp{\begin{split}}
\def\ensp{\end{split}}

\begin{document} 

\author{M. Cosentino Lagomarsino}
\affiliation{UMR 168 / Institut Curie, 26 rue d'Ulm 75005 Paris, France}
\email[ e-mail address: ]{mcl@curie.fr}
\author{P. Jona} 
\affiliation{Politecnico di Milano, Dip. Fisica, Pza Leonardo Da Vinci
  32, 20133 Milano, Italy} 
\email[ e-mail address: ]{ patrizia.jona@fisi.polimi.it}
\author{B.  Bassetti} 
\affiliation{Universit\`a degli Studi di Milano, Dip.
    Fisica, and I.N.F.N. Via Celoria 16, 20133 Milano, Italy } 
\email[e-mail address: ]{ bassetti@mi.infn.it }

\title{The large-scale logico-chemical structure of a transcriptional
  regulation network}

\begin{abstract}
  Identity, response to external stimuli, and spatial architecture of a living
  system are central topics of molecular biology. Presently, they are largely
  seen as a result of the interplay between a gene repertoire and the
  regulatory machinery of the cell.  At the transcriptional level, the
  \emph{cis}-regulatory regions establish sets of interdependencies between
  transcription factors and genes, including other transcription factors.
  These ``transcription networks'' are too large to be approached globally
  with a detailed dynamical model.  In this paper, we describe an approach to
  this problem that focuses solely on the \emph{compatibility} between gene
  expression patterns and signal integration functions, discussing
  calculations carried on the simplest, Boolean, realization of the model, and
  a first application to experimental data sets .
\end{abstract}

\maketitle

\section{Introduction}


Regulation can be defined as the set of physico-chemical constraints operating
within a living cell that modulate the expression of the cell's genes. In the
present view of molecular biology, regulatory processes are often used as a
primary causal explanation for many phenomena, playing a role in this
discipline that is comparable to the role fundamental interactions play in
physics.  In fact, it is widely believed that the repertoire of signal
responses (and, more in general, of all the information processing and
structural tasks) of living systems is encoded in interconnected threads of
genes regulating the activity of each other.  These networks of
interdependencies are still largely uncharacterized, although they have begun
to fall within reach of systematic experimentation in the recent
years~\cite{HCP04,NRF04,BLA+04,WA03,LRR+02}.

Considering the so-called ``central dogma'' of molecular biology, 
\begin{displaymath}
\textrm{DNA} \stackrel{\textrm{transcription}}{\longrightarrow}
\textrm{mRNA} \stackrel{\textrm{traslation}}{\longrightarrow}
\textrm{protein} \stackrel{\textrm{folding}}{\longrightarrow}
\textrm{function}, 
\end{displaymath}
regulation processes can intervene at all the separate steps (and also in
different sub-steps).  Regulation exploiting the process of transcription, or
transcriptional regulation, constitute to date the best understood among all
the possible regulation mechanisms.

\begin{figure}[htbp]
  \centering
  \includegraphics[width=.85\textwidth]{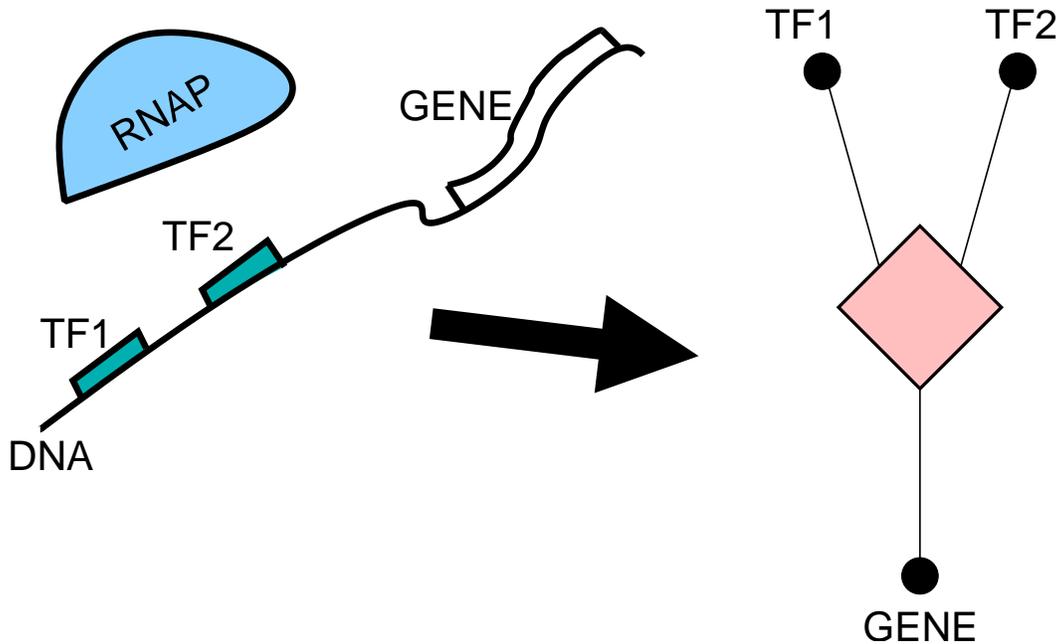}
  \caption{Schematics of our representation of a signal integration function
    at the \emph{cis-}regulatory region of a gene as a constraint on the gene
    expression variables. For general variables, the constraint involves
    minimization of the free energy of the Shea-Ackers model. In GR1, the
    constraint is Boolean.}
  \label{fig:Fagraph}
\end{figure}

\begin{figure}[htbp]
  \centering
  \includegraphics[width=.8\textwidth]{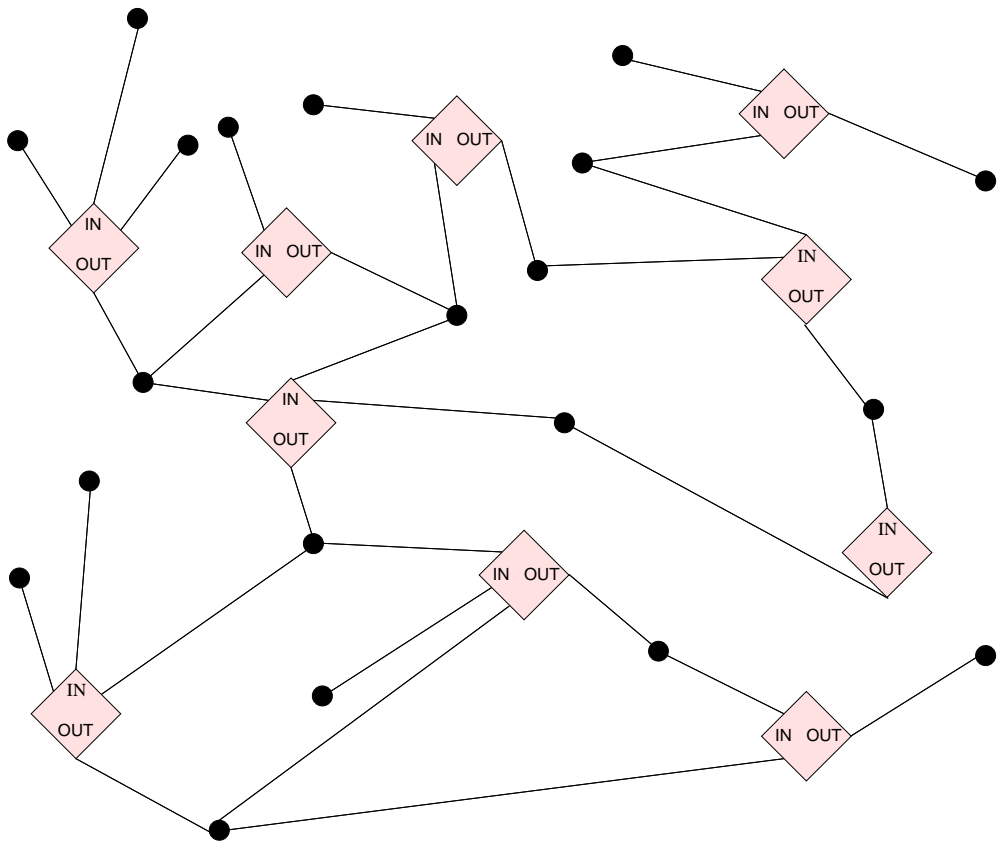}
  \caption{Graph representation of a transcription network.  Each diamond node
    represents a signal integration function, while each black circle is a
    variable. The directionality of the constraint is represented graphically
    by labeling the diamonds with IN and OUT on two different sides.}
  \label{fig:grafone}
\end{figure}

Transcriptional regulation networks are defined starting from the basic
functional building blocks involved in transcription. These are (i) the
promoter region of a gene or operon along the DNA sequence, which contains the
\emph{cis} regulatory binding sites for the transcription factors, (ii) the
transcription factors, which are proteins that regulate the binding of
RNA-polymerase, and (iii) RNA-polymerase, the protein complex that performs
transcription of a gene or an operon in mRNA form~\cite{Pta92,ABL+03}.  The
amount of mRNA transcribed is related to the expression of a particular gene
only if one takes for granted all the other steps that bring to a functional
protein. If this (big) leap is accepted, the ``state'' of a cell is identified
to the mRNA concentration of its genes. Experimentally, this is particularly
sound for prokaryotes and simple unicellular organisms, but often assumed in
more complex contexts, for example in DNA microarray experiments.
Under this assumption, the locations and orientations of the binding sites for
transcription factors, as well as the affinity of the transcription factors to
different binding sites, determine the expression levels of a gene in response
to changes in the active transcription factor concentrations inside the cell.
In turn, the concentration of active transcription factors (the ones that can
actually bind) encodes the configuration of the environment, for example
through degradation or activation by internal and external signaling
molecules.
A \emph{cis-}regulatory region can contain many binding sites for many
transcription factors which act in cooperation (or competition) on the
promoter region, to control in a combinatorial way the binding of RNA
polymerase.  This process, referred to as signal integration, is the
logic heart of the network. 
%
A transcriptional regulation network can be represented as a hypergraph
containing both gene expression (``variable'') nodes and signal integration
(``function'') nodes. The connectivity is the source of the network complexity
(Fig.~\ref{fig:grafone}).

Thus, a transcriptional regulation network can work independently as a
computational unit in a living cell, being able to make decisions on which
genes will be switched on at different times. Studies focusing simply on the
\emph{structure} of the underlying graph have lead to interesting
results~\cite{BLA+04,SMM+02,WtW04}.  However, characterizing and predicting
gene expression patterns given a network structure remains an enormous
challenge. Two main problems exist.  Firstly, the networks are only partially
characterized experimentally~\cite{HCP04,NRF04}.  In some
instances~\cite{BLA+04,DRO+02}, the wiring diagram is well known, but in
general the functions are described only qualitatively, typically with
annotations such as activation, repression or dual effects, and little is
known about their actual structure.
Secondly, transcription networks are fairly large. While detailed models or
simulations work well on small (sub-)systems~\cite{MA97,ARM98}, typically a
coarse grained approach is needed.
%
Microscopically, it is well accepted that the Gillespie algorithm~\cite{GD77},
while disregarding spatial correlations, correctly describes the stochastic
asynchronous events of chemical kinetics involved. On the other hand, with a
mesoscopic average in time, it is still unclear what the emergent time scales
might be.  In particular, the pioneering approach of
Kauffman~\cite{Kau69,Kau69b,Kau93,Kau04}, suggesting a synchronous
deterministic dynamics for a Boolean (i.e.  ON/OFF) representation of the
network is still being debated, both in its assumptions and in its
results~\cite{Ger04,KD05,SaK03,BP98}.


We consider the second problem, and develop a model (called GR, from Gene
Regulation), that focuses, rather than on pure dynamics, on the compatibility
between gene expression patterns and signal integration functions. The
compatibility constraints are generated by the clauses encoded by signal
integration functions at \emph{cis}-regulatory regions.  Our framework
describes the system as a combinatorial optimization problem where $N$
variables, the gene expression levels, are subject to $M$ constraints,
representing the signal integration nodes.  Simply put, a cell with $N$ genes
can express them in exponentially many ways, $2^N$ in the Boolean ON/OFF
representation. However, the cell never explores all the possible patterns of
expression. It generates only clusters of correlated configurations.  To fix
the ideas, one can think to the very elementary example of the cI-cro switch
of $\lambda$-phage~\cite{Pta92,Tho73}. In this case one could observe the
states 10 (where cI is ON, and cro is OFF), 01, or perhaps 00, but never 11,
because this state is ruled out by the signal integration function.
In a cell, with the added complexity of the regulation network, we can think
that many of the states are not observable for the same compatibility reasons.
The approach is easily connected to detailed thermodynamic treatment of
transcription from a
signal integration node on one side~\cite{BGH03}, and to the statistical
mechanics of spin glasses and combinatorial optimization problems on the
other~\cite{MPZ02}.

%
Rather than the detailed quantitative prediction of mRNA expression states,
the current challenge is to set a conceptual framework which can help to
interpret the observations in concrete examples, integrating as much as
possible with known data.
As a the simplest example of this, we study the behavior of the Boolean
version of our model on the network structure of E.~coli.
Building up from this simplest case, the aim is to analyze increasingly
realistic network structures, in order to generate a theory that, while being
consistent with the generic qualitative features of regulation networks, is
useful to analyze single instances and realizations. This is maximally
important as biological knowledge is constructed on specificities, and not on
typical case behavior.


This paper is structured as follows. Sec.~\ref{sec:model} introduces the model
abstractly, as an optimization problem, which, in sec.~\ref{sec:shea} is
connected to the more concrete thermodynamic Shea-Ackers model of
transcriptional regulation. Sec.~\ref{sec:satmap} abandons this general
setting, and takes on the simplest possible formulation of the model, GR1,
which has Boolean functions and variables, showing that this case maps
directly to a so-called satisfiability problem (Sat). The scope of
sec.~\ref{sec:leaf}, is to analyze the typical number $\mathcal{N}$ of gene
patterns of a random instances of GR1, starting from the case of fixed
connectivity. The ``leaf removal'' algorithm allows to carry this analysis in
the annealed approximation.  An important premise is the fact, which is
evident looking at the data~\cite{SMM+02}, that some genes are essentially
``free'' from the point of view of transcription. These are mainly controllers
and are connected to external stimuli. The expression of the rest is
conditioned to the state of other genes.  The algorithm allows to define the
``complex combinatorial core'' (CCC) of the network, as the set of genes able
to control its global state.  The number of non-controlled, or ``free''
variables in the core determines the complexity of the system. The phase
diagram shows three distinct regimes of gene control.  In the first (UNSAT),
there are no free genes in the core, and the system cannot control the
simultaneous expression of all its genes.  In the second regime (``complex
control'' or HARD-SAT), the core contains free genes that control, both
directly and indirectly, many others.  The general dynamics is residual (many
variables are fixed, the others can change).  In the third regime, the core is
empty.  Each free gene (which is external to the core) controls the state of a
small number of genes (``simple control'', or SAT phase).
Sec.~\ref{sec:selfav} concerns itself with the \emph{width} of the
distribution of $\mathcal{N}$, which has both a technical significance as a
validity test of the annealed approximation, and a biological one, as the
variability in the number of gene patterns at fixed gene number.
Sec.~\ref{sec:multipoiss} discusses generalizations of these results to
non-fixed connectivities.  Finally, sec.~\ref{sec:concr} describes one first
attempt to put this findings to work on an experimental data set.

\section{Model}
\label{sec:model}

Our aim is to describe in a minimal way gene expression in a transcription
network, separating the issues related to the dynamics from those related to
its logical and computational structure. In order to do this, we will
formulate a model that sees the system as an optimization problem, where a set
of variables, the genes, is subject to a set of constraints, the signal
integration nodes. Upon this logic backbone, many a dynamics can be
superimposed, including in the most general case the kinetic Montecarlo scheme
commonly used to model genetic networks.
Rather than going towards the direction of highest detail, we will choose to
simplify the model as much as possible, reducing the number of details to the
minimum, and studying the general qualitative features of the system.

The model is specified by 
\begin{itemize}
\item[1)] A set of $N$ discrete variables $\{ x_i \}_{i=1..N}$ associated to
  genes or operons, which in the simplest picture are identified with their
  transcripts and protein products. These variables represent the expression
  levels and in general take discrete values in $\{0,..,q\}$. In particular
  situations, they are well-approximated by continuous variables.
\item[2)] A set of $M$ interactions, or constraints $\{ I_b(x_{i_{0}},
  x_{i_1}, ..., x_{i_{k_b}}) \}_{b=1..M}$ between the genes, representing the
  signal integration from transcription nodes.
\end{itemize}
This formulates an optimization problem, which we call GR, from Gene
Regulation. GR asks to find the states compatible with the constraints.

The model can be easily generalized to include other relevant degrees of
freedom, such as translation, protein modification and protein-protein
interactions. However, each addition adds complexity and parameters.
Therefore we start with the minimal possible description.
Admittedly, neglecting non-transcriptional regulation is a drastic
simplification of the system. A complete genetic network should in principle
include all forms. On the other hand, the justification for considering
transcription alone is that it is the first step in the chain of regulation
events and it is experimentally well characterized.  From the physics point of
view, this model can be seen as a ``spin glass'', a system where some
variables, our gene expression levels, interact through some coupling
constants, specified by the constraints~\cite{M02}. This approach to
optimization problems of computer science has proved to be very useful in the
recent years~\cite{MPZ02}.

The network structure is naturally represented on a ``factor graph'', where
two kinds of nodes are present, N ``variable nodes'' and M ``function nodes''
respectively (Fig.~\ref{fig:grafone}). Here $k_{b} = 1+ K_{b}$ can be seen as
the local connectivity of a function node. Note that the factor graph is also
defined by a variable connectivity $c_{i}$, the number of functions connected
to $x_{i}$.
In fact, this is the typical structure of a constraint satisfaction problem of
theoretical computer science, such as q-coloring or satisfiability~\cite{M02}.



\section{The constraints and the Shea-Ackers model}
\label{sec:shea}

To specify the model one has to give a structure for the function nodes, i.e.
the constraints. This requires a physical model for signal integration. In
order to take this step, in this section we start from the well-known and
widely accepted thermodynamic model of Shea and Ackers of gene activation by
recruitment, to show that it is the natural setting to express
our constraints.  

\begin{figure}[htbp]
  \centering
    \includegraphics[width=.7\textwidth]{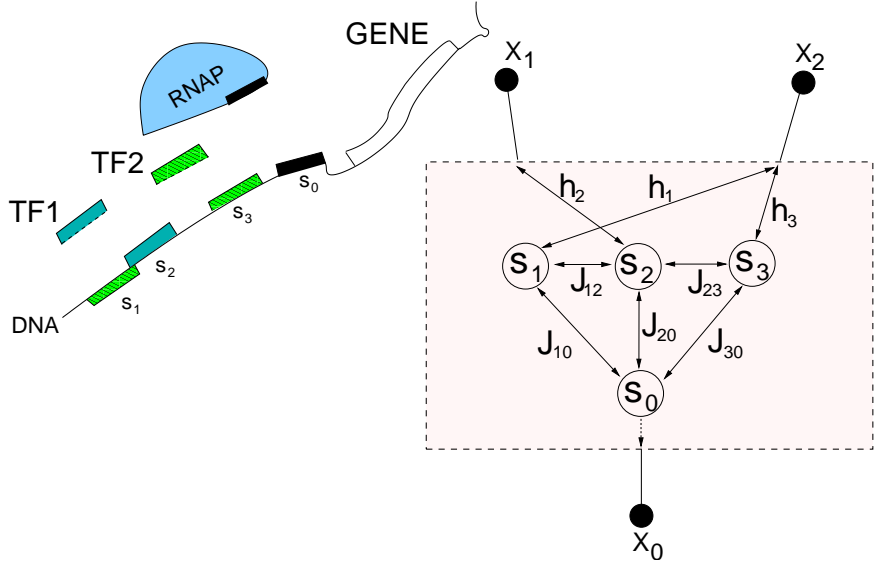}
  \caption{Exemplification of a Shea-Ackers node. $s_i$ are occupation
    variables for the transcription factors binding sites, while the coupling
    constants $J_{ij}$ encode cooperative or competitive binding. The external
    fields $h_i$ are the phase space variables of GR.}
  \label{fig:Sheanode}
\end{figure}

Let us consider a function node with $K$ regulators and one output variable.
This is modeled, in the version presented by Buchler and
collaborators~\cite{BGH03}, as a neural network (a ``Boltzmann machine'') with
Hamiltonian
\begin{displaymath}
  H = \sum_{\stackrel{i,j = 0..L}{ i\ne j} } J_{ij} s_i s_j
  + \sum_{j = 0..L} h_j s_j \ \ ,  
\end{displaymath}
where $s_1, .., s_L$ are the occupation variables of the \emph{cis-} binding
sites, $h_{j}$ are external fields, functions of $x_1, ..., x_K$ representing
the concentrations of ``input'' transcription factors, and $J_{ij}$ are
interaction constants associated to competitive versus cooperative binding.
More precisely, $h_{j} = - \beta^{-1} \log(Q_{j})$, where $Q_{j} =
\frac{[TF_{i}]}{\kappa_{i}} \sim \frac{x_{i}}{\kappa_{i}}$ is the binding
affinity of a site $i$, and $\kappa_{i} $ a dissociation constant. Normally,
the concentrations are approximated with continuous variables.
In general, $L \ge K$, because multiple binding sites are present.  Finally,
$s_0, h_0$ are the occupation variable and the external field (a fixed
parameter corresponding to the polymerase binding affinity) associated to the
output node of the function (see Fig~\ref{fig:Sheanode}).

Given all the binding constants and the interaction parameters, the
-intrinsically probabilistic- output of the gate is computed as a function of
the input fields, simply as the probability that $s_0 = 1$.
This expectation value can be obtained through the partition function
\begin{displaymath}
  Z[h_1,...,h_L] = \sum_{\{s\}} e^{- \beta H}
\end{displaymath}
as
\begin{displaymath}
  P(\sigma_0 = 1) = \frac{1}{Z}  \sum_{\{s\}}  e^{- \beta
  H(1,s_1,...,s_L)} 
\end{displaymath}
where $\beta = 1/kT$. 

Equivalently, one can compute the local free energy
\begin{equation}
  - \frac{1}{\beta} \log Z = F[h_0,...,h_L]  = F[x_0, ..., x_K]
  \label{eq:fensa}
\end{equation}
and find the average output through minimization with respect
to the $x_0$ coordinate.

In other words, the function nodes are local equilibrium conditions for the
variable nodes, specified by the Shea-Ackers model of the
\emph{cis-}regulatory region of each variable node. The expression variables
$\{ x_i \}_{i=1..N}$ need to satisfy the constraints specified by the local
minimizations of the free energies $\{ F_b(x_{i_{0}}, x_{i_1}, ...,
x_{i_{k_b}}) \}_{b=1..M}$. Since there is a clear input-output logic encoded by
the chemical equilibrium of each signal integration nodes, one could refer to
this backbone static structure the ``logico-chemical'' structure of the
network, and separate it from its ``dynamic'' structure.  The logic it encodes
is of course not Boolean. In fact, it is intrinsically non-Boolean even with
Boolean variables, as the outputs are probabilistic functions of the inputs.

From the point of view of statistical mechanics, this is a Potts spin system
with diluted interactions described by the local free energies $F_b$ (which in
this context should be interpreted as effective Hamiltonians).
Interestingly for the analogy with spin glasses, the model for the gate can be
seen as a message-passing procedure analogous to that exploited by the cavity
method~\cite{MPV87,MP03,MZ02}, where, in the approximation of factorized
probability distribution of the variables, one evaluates the local fields
$h_{i \rightarrow b}$, describing the local influence of the couplings on
variable $i$ in absence of interaction $b$, and $u_{b \rightarrow i}$, the
contribution of interaction $b$ on the local magnetic field on spin $i$,
together with their histograms in the presence of many states.
In our case, the ``messages'' described above can only travel in the
input-to-output direction.  We are currently investigating whether this
analogy can be exploited for further calculations, and we are aware of work in
this direction, in a simpler setting, by another group~\cite{CLP+}.

In principle, the variables $x_i$ directly stand for the number expressed of
molecules in a cell. Provided the set of all binding constants and
interactions is known, all the function nodes can be computed and the model is
complete.  It could be solved, for example by numerical simulations, once a
dynamics is specified. 

On the other hand, with a few exceptions of small systems, these (many)
parameters are in general not known. For this reason, rather than
aiming for the highest level of detail, we choose to simplify as much
as possible, while trying to keep the most relevant features.
For practical purposes, in order to be able to advance further analytically
and numerically in the understanding of the model, it is convenient to
introduce coarse grained expression levels, thereby effectively reducing $q$.
The resulting model, GRq, is identical, mind the fact that variables and
constraints are now subject to implicit averaging. One advantage of this
approach is that local free energies become easier and easier to specify, and
it is possible to study, as is commonly done for spin glasses, the typical
behavior of the system as a function of the parameters.

In the simplest possible scenario $q=1$, and the expression levels are Boolean
variables. The assumption behind this is that what matters is only if the
level of expression is high or low~\cite{Kau93,Kau69b}.
The simplest case (which we will still call GR1), assumes also Boolean
functions.  In the following section, we will show how GR1 maps to a
Satisfiability problem (Sat), an optimization problem where $N$
Boolean variables are constrained by $M$ conjunctive normal form (CNF)
constraints (i.e. by a Boolean polynomial constructed as a product ($\wedge$)
of $M$ disjunctive monomials ($\vee$)).

\section{GR1, mapping on a Satisfiability Problem}
\label{sec:satmap}

As we have shown in the above section, for general variables $x_i$ that
represent real expression levels, the constraints can be derived directly from
the model of Shea and Ackers of gene activation by
recruitment~\cite{SA85,BGH03}. If the $x_i$ represent coarse-grained
expression levels, the same model can be used to construct the local free
energy in Eq.~\eqref{eq:fensa}, associated to each signal integration node,
that generates the constraints through minimization.
Here we consider the simplest possible scenario, treating the
expression levels as Boolean variables, setting $q=1$, and the signal
integration functions as boolean functions $\{ f_b(x_{b_1}, ..,
x_{b_{k_b}}) \}_{b=1..M}$. 
We also restrict to the case of fixed in-ward connectivity $k_b = K, \ \ 
\forall b$.  These conditions, defining $K$-GR1, are also found in Kauffman
networks~\cite{Kau93,Ger04}. We will relax the hypothesis of fixed $K$ to
explore networks with fluctuating connectivity in section
\ref{sec:multipoiss}.

\begin{figure}[htbp]
  \includegraphics[width=.7\textwidth]{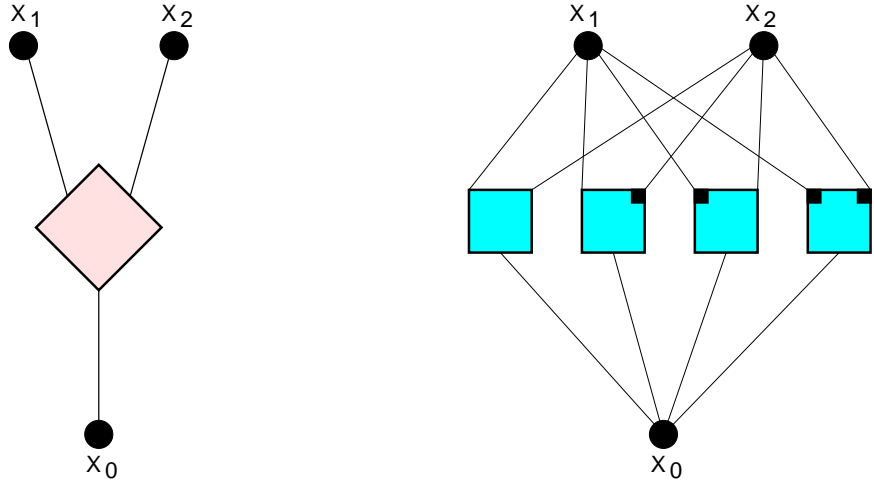}
  \caption{Translation of a 2-GR1 node in 3-Sat Nodes. The
    input-output direction is from top to bottom. Sat constraints are
    represented as squares, where the black and white vertices indicate that
    the corresponding variable enters negated or affirmed respectively in the
    3-Sat constraint.}
  \label{fig:schemino}
\end{figure}

The expression
\begin{math}
   x_{b_0} = f_b (x_{i_1}, .. ,x_{i_{K}}),
\end{math}
which imposes that the variable $x_{b_0}$ is the output of the
function $f_b$, translates into the Boolean constraint
\begin{equation}
  \neg(x_{b_0} \dot{\vee} f_b).
  \label{eq:constr}
\end{equation}
In a Kauffman network, this expression is equivalent to the fixed
point condition, and there is one such constraint for every variable
$x_i$. The formula
\begin{displaymath}
  I = \bigwedge_{b=1..M} I_b = \bigwedge_{b=1..M} \neg(x_{b_0} \dot{\vee} f_b)
\end{displaymath}
defines a Satisfiability problem (Sat) on the variables $x_1,.., x_N$.
From the biological viewpoint, this is a logic representation of the
computational tasks encoded in the transcription network by evolution,
i.e. which sets of genes have to be switched on at any given
condition.  More abstractly, having in mind Kauffman networks, each
satisfying solution of this problem corresponds to a fixed point in
the Kauffman dynamics (independently from the update scheme).  The
number of variables involved in one constraint is always exactly
$k=K+1$, therefore this mapping associates a network with fixed
connectivity $K$, to a $k$-Satisfiability problem whose connectivity is
increased by one unit.  For example, a $K=2$ Kauffman network
corresponds to a 3-Sat problem, and so on.
The suitable order parameter for such a system is $\gamma = M/N$. GR1 assumes
that each gene expression variable is regulated at most by one signal
integration function, so that $\gamma \le 1$.

To further understand the logic structure if GR1, we can write the CNF
constraints on each variable $x_n$. This allows to make a connection to the
order parameters used in $k$-Sat, i.e. the local constraint $\alpha$, defined
as the number of conjunctive-normal-form (CNF) clauses per Boolean variable.
In order to do this, we recast the Boolean formulas $I_n$ into CNF.
Reshuffling the truth table of $f_n$ in a way that the first $z$ terms
($1,..,z$) give zero as an output, a simple procedure shows that
\begin{equation}
\begin{array}{ccc}
  I_b &=& 
  \Big( \bigwedge_{\alpha=1}^{z} 
    (\neg x_{b_0} \vee \xi_{\alpha_1} \vee ..\vee \xi_{\alpha_K})   
  \Big) \\ & & \\
 &&  \bigwedge 
  \Big(
    \bigwedge_{\alpha=z+1}^{2^K} ( x_{b_0} \vee \xi_{\alpha_1} \vee ..\vee 
    \xi_{\alpha_K})  
  \Big),
\end{array}
   \label{eq:satform}
\end{equation}
with
\begin{displaymath}
\xi_{\alpha_j} =
\left\{
  \begin{array}{cc}
    x_j & \textrm{if element } \  \alpha,j \ \textrm{of truth table} = 0 \\
    \neg x_j & \textrm{if element } \ \alpha,j \ \textrm{of truth table} = 1
  \end{array}
\right. ,
\end{displaymath}
having exactly $2^K$ clauses of $K+1$ elements.  
Thus, a network with connectivity $K$ maps into a $(K+1)$-Sat having always
$\alpha = 2^K \gamma$.  
However, we cannot imply directly that the typical behavior, and therefore the
phase diagram of the system will be the same as that of the corresponding
random $k$-Sat. Considering random realizations of the constraints, these are
\emph{a priori} only a subset of all the possible realizations of a $k$-Sat
constraint. In fact, it is immediate to realize that the $2^K$ CNF clauses
written in Eq.~\eqref{eq:satform} contain all the possible (fixed)
combinations for the inputs variables and the corresponding (random) $2^K$
outputs for the output variable. This is best exemplified on the factor graph
(Fig.~\ref{fig:schemino}).

Having established a link between GR1 and a particular optimization problem we
set out to look at random instances for the signal integration functions. For
fixed connectivity one can expect a similar behavior for $K$-GR1 as random
$k$-Sat, or $k$-XORSAT (a Sat problem with clauses containing only XOR
clauses), with the presence of a phase transition in the number of satisfying
states.  The suitable order parameter for such a transition is $\gamma = M/N$.
Notably, the space of functions of the three models have different dimensions.
Furthermore, differently from Sat or XORSAT, in GR1 each gene expression
variable is regulated at most by one signal integration function, so that
$\gamma \le 1$. In practice, both for ease of interpretation and for
simplicity of the analytical formulation, from now on we will abandon the
formulation of GR1 in terms of CNF constraints, and work with the input-output
functions $f_{n}$.

\section{Leaf Removal and the Computational Core}
\label{sec:leaf}

The aim of this section is to compute the number of satisfying solutions
$\mathcal{N}$, for random instances of the constraints. For accessory reasons,
we will map GR1 to a spin system.  Quite simply, each Boolean variable $x_i
\in \{0,1 \}$ is transformed into a spin $\sigma_i \in \{-1,1\}$.
Each constraint, or diamond function node generates the interaction
Hamiltonian $ H_{\diamond,b}$
\begin{equation}
  2^k H_{\diamond,b} = \sum_{J_{b_1},..,J_{b_K}} 
                       \prod_{l=1..K} (1+ J_{b_l} \sigma_{b_l}) 
                       (1+ J_{b_0}^{\{ J_{b_1},..,J_{b_K}\}} \sigma_{b_0})\ ,
   \label{eq:ham_diam}
\end{equation}
Under the restrictions defining GR1, the total energy of the system is simply
the number of violated constraints.  A zero-energy configuration satisfies all
the constraints, and is therefore able to comply to all the logic functions
encoded by the network.  The $2^K$ coupling constants $J_{b_0}^{\{
  J_{b_1},..,J_{b_K}\}} = \pm 1$ are a representation of the truth table of
the function $f_b$.  With the correspondence $\{0,1\} \leftrightarrow
\{-1,1\}$ , $J_{b_1},..,J_{b_K}$ stand for the possible values of the input
variables $ x_{b_1},..,x_{b_K}$, while $J_{b_0}^{\{J_{b_1},..,J_{b_K}\}}$ is
the associated output value.  The Hamiltonian (\ref{eq:ham_diam}) is the cost
function of the corresponding optimization problem K-GR1. It encodes the logic
constraint of Eq.~(\ref{eq:constr}), in the sense that it is one whenever the
constraint is violated, and zero otherwise.  From the point of view of
transcription networks, $H_{\diamond,b}$ is the coarse grained local free
energy of the Shea-Ackers model. Note that, even in the case of Boolean
variables, the coupling constants represent binding affinities and
interactions between transcription factors. In general, they need not be plus
or minus one. Here we took this further assumption. 

The conventional average of $\mathcal{N}$ on the realizations might be biased
by the weight of exceptions~\cite{MPV87}. The correct quantity to compute is
the ``quenched average'' of the system's free energy,
$\overline{\log{\mathcal{N}}}$, which is usually accessed with the replica, or
similar methods~\cite{MPZ02}, passing from Hamiltonians like $H_{\diamond,b}$.
%
%
However, in the case under examination we will use a simpler method, based on
the ``leaf removal''~\cite{MRZ03} algorithm, which allows to compute only the
\emph{annealed} average $\log{\overline{\mathcal{N}}}$. As we will discuss,
this method has the advantage of an immediate biological interpretation in
terms of the roles played by genes in the network.
For the case of random XORSAT, Mezard and collaborators~\cite{MRZ03} have
shown that the annealed average on the core variables coincides with the
quenched one. In general $\overline{\log{\mathcal{N}}} \leq
\log{\overline{\mathcal{N}}}$.  
For GR1, we performed estimates that indicate the presence of the same
self-averaging property in a well-defined region of parameter-space (see
sec.~\ref{sec:selfav}). Within this region, our annealed calculation is exact.

For a given realization of the constraints $ \{ \vec{I}, \vec{f} \} $, the
number of satisfying states $\mathcal{N}$ can be written as
\begin{displaymath}
  \mathcal{N}(\vec{I}, \vec{f}) = \sum_{\vec{\sigma}} \prod_{b=1}^M \delta(1;
  f_{b}(\sigma_{i(b,1)}, .., \sigma_{i(b,K_b)})\sigma_{i(b,0)}). 
\end{displaymath}
Here, the randomness is contained: (i) in the specification of the network
structure, $\vec{I} = (I_1,...,I_M)$, i.e. in the coordinates $i(b,l)$, which
point at the variable occupying place $l$ in the $b$th constraint; (ii) in the
specification of the functions $\vec{f}= (f_1,...,f_M) $ with a certain
probability distribution in the class $\mathcal{F}$. An overbar ($\bar{\ ~}$)
indicates an average on both distributions, $p(\vec{I})$ and $p(\vec{f})$.  We
will first concentrate on the case with fixed in-ward connectivity $K$.

In carrying this average, there are three relevant preambles. The first is
that not all the $M$ equations and $N$ variables are meaningful to calculate
$\mathcal{N}$.  Indeed, every output variable that appears in only one
constraint can be trivially fixed according to its function. Thus, both the
constraint and the variable can be eliminated without affecting the number of
solutions.  This procedure is called ``leaf removal''~\cite{MRZ03}.  It is a
nonlinear procedure, as more variables can disappear together with a single
constraint, because input free genes that regulate a leaf remain as isolated
points.  The iteration of this mechanism leads to the definition of a
``core'', the CCC, of significant variables and constraints, in numbers of
$N_C$ and $M_C$ respectively.  In the CCC, $M_{C}$ genes are controlled, and
$\Delta_{C} = N_{C} - M_{C}$ are the ``free'' genes with an essential role in
controlling the expression states, as a function of an input signal.
The second relevant fact is the hypothesis that the functions are independent
and identically distributed random variables.  Thirdly, we consider a set of
functions in a family $\mathcal{F}_K$ which satisfy the condition
\begin{math}
  \frac{1}{2^{2^K}} \sum_{f \in \mathcal{F}_K } p(\vec{f}) f(\vec{x}) = \rho
  \, ,
\end{math}
as will always be the case if the outputs of the functions are
uncorrelated, even in the presence of bias. 

It is then easy to verify that
\begin{displaymath}
  \overline{ \mathcal{N}} = \sum_{\vec{\sigma}, \vec{I}} p(\vec{I}) 
  \prod_{b=1}^M \left(
    \rho \delta(1;\sigma_{i(b,0)}) + (1- \rho) \delta(-1;\sigma_{i(b,0)} 
    ) \right) \ ; 
\end{displaymath}
so that, as for the XORSAT model, 
\begin{equation}
  \overline{\mathcal{N}} = 2^{N_{C}-M_{C}} \ . 
  \label{eq:nstati}
\end{equation}
Incidentally, we note that the same procedure is valid for GRq, where
$\overline{\mathcal{N}} = q^{N_{C}-M_{C}}$.

\begin{figure}[htb]
  \includegraphics[width=.75\textwidth]{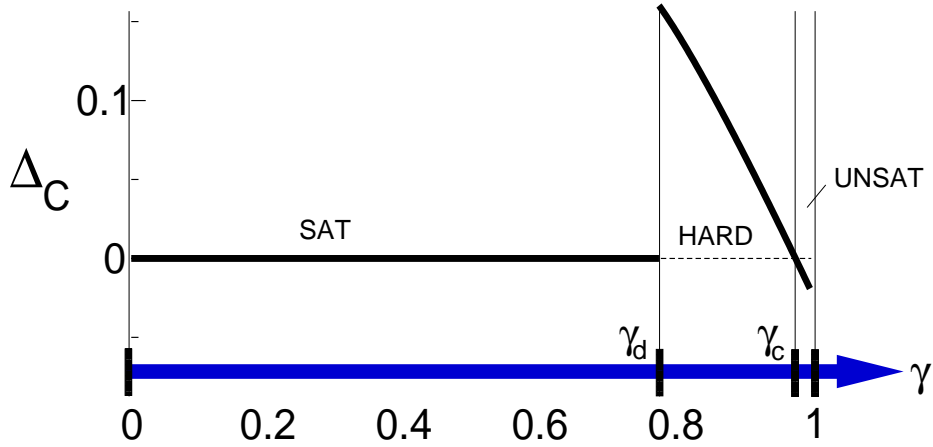}
  \caption{Phase diagram of $4$-GR1. For $\gamma >
    \gamma_c$ no solutions exist in the typical realization (UNSAT phase). For
    $\gamma < \gamma_d$ the system is paramagnetic typically a satisfying
    state exists (SAT phase, or simple gene control). For $\gamma_d <\gamma<
    \gamma_c$ there is complex gene control. }
  \label{fig:pd}
\end{figure}

We can use $\frac{\Delta_C}{N_C}$ as an order parameter in the thermodynamic
limit $N \to \infty, \, \gamma$ const., to distinguish three types of
phenomenology, or three ``phases''. (i) For $\frac{\Delta_C}{N_C} \leq 0$,
there are no free genes in the core, and the system cannot comply to all the
expression programs encoded into its DNA. This is a UNSAT phase from the
computational point of view. (ii) For $0 \leq \frac{\Delta_C}{N_C} < 1$, the
$\Delta_{C}$ genes of the CCC control $O(N)$ genes each, and therefore are
able to determine an expression state. This is a ``complex gene control''
phase, or HARD-SAT (iii) For $ \frac{\Delta_C}{N_C} = 1$ the number of
controlled variables is a vanishing fraction of the total number of genes.  In
other words, $ M_C = 0$ in the thermodynamic limit, the free genes control at
most $O(1)$ genes to generate a satisfying state (``simple control'', or SAT
phase).  In the simple control phase, the system is underconstrained, which
means that the logic conditions imposed by the signal integration functions
are generally insufficient for a strict determination of the expression
patterns.

The three phases described above depend both on the value of $\gamma$ and on
the class of random functions considered. In general, if all the possible
functions are taken into account, the phase diagram can be explored studying
the rank and the kernel of the connectivity matrix~\cite{CS02}.
%
%
%
%
Following the analysis of Mezard and collaborators~\cite{MRZ03}, in the case
of Poisson variable connectivity, i.e. with the distribution $\pi(c) =
\frac{(k \gamma)^{c}}{c!} e^{-k \gamma} $, the phase diagram as a function of
$\gamma$ is identical to the random XORSAT problem.  It is illustrated in
Fig.~\ref{fig:pd}. For $\gamma>\gamma_c$ no solutions exist in the typical
realization (UNSAT phase).  For $\gamma<\gamma_d$ the system is paramagnetic
(SAT phase). For $\gamma_d<\gamma<\gamma_c$ exponentially many satisfying
states exist.
Here, the space of solutions breaks down into clusters separated by free
energy barriers.  The typical dynamics in a cluster will be residual, in the
sense that a block of genes are fixed (on or off) and the rest may move.  The
number of clusters is controlled by the (computational) complexity $\Sigma$ of
the system.  The number of observed configurations is $ \mathcal{N^{*}} \sim
\exp[N \Delta_{C}]$.  Thus, by definition $\Sigma$ is directly related to the
order parameter $\Delta_{C}/N_{C}$, i.e. to the partitioning of the core
genes.  How the system explores (or not) the clusters depends on details of
its dynamics.







\section{State-Fluctuations and Self-average}
\label{sec:selfav}

In this section, we discuss a calculation of the \emph{width} of the
distribution of the number of compatible states.  In order to do this, we
compute the quantity $\overline{[\mathcal{N}]^2}$.  This calculation is
relevant both from the technical and from the qualitative point of view. The
technical aspect, as already anticipated, deals with the self-averaging
property, which holds when the quantity
\begin{displaymath}
  \frac{\overline{[\mathcal{N}]^{2}} - \left[\overline{\mathcal{N}}\right]^2}
  {\left[\overline{\mathcal{N}}\right]^2}  
\end{displaymath}
vanishes in the thermodynamic limit $N \to \infty$ at constant $\gamma$.  When
this condition is met, the annealed average computed above coincides with the
quenched one, and no extra effort is required. When it is not, the behavior of
the system can be qualitatively different from what emerges in the annealed
picture. In particular, the typical number of solution is overestimated.  In
this case, more complicated formalisms, such as replicas, need to be
adopted~\cite{MPV87}.

The qualitative aspect involves the possible physical, and biological,
interpretation of $\overline{[\mathcal{N}]^2}$. This is an indicator of the
width of the probability distribution in the number of compatible states, in
presence of random functions and network structure.  Therefore, it can be seen
as the freedom the system has of varying the number of states that comply to
the signal integration functions by acting on its constraints.  Biologically,
having in mind Darwinian evolution one may interpret it as a kind of
``adaptability''. This is done as follows. If $\mathcal{N}$ is interpreted as
the number of possible responses of gene patterns to external or internal
conditions, it is reasonable to assume that a given system with a fixed number
of genes will be fitter by maximizing $\mathcal{N}$. Now, if the distribution
is wide, the system can vary greatly $\mathcal{N}$ by acting on the signal
integration functions.  If the distribution is peaked, the changes in the
functions will be less effective in increasing the number of states. 

If the self-averaging property holds, all the width, this adaptability, comes
only as a finite-size effect. This is not irrelevant, considering that the
value of $N$ is in the range $10^{3}-10^{5}$, quite far from the usual
Avogadro's number!
On the other hand, if the self-averaging property does not hold, it means that
a residual width exists even in the thermodynamic limit. In this case, an
evolved system can be ultra-specific, finding the very exceptional situations
in which a high number of solutions exists against the typical odds in which
the system cannot express compatible gene patterns, the biblical needle in a
haystack. Such putative highly specialized organisms would be particularly
sensitive to changes in the environment. Given these considerations, it seems
that the lack of self-averaging for a model like GR1 would make it slightly
less appealing.

The final result for GR1 is that $\mathcal{N}$ is indeed self-averaging.  The
details of the calculation are reported in Appendix~\ref{sec:self-aver-prop}.
It relies on two basic assumptions. The first relates to the choice of a
family of random functions such that: 

(a) 
\begin{math}
  \frac{1}{\Lambda} \sum_{f \in \mathcal{F}} p(\vec{f})
  f(\vec{\s}) = 0 \, ,
\end{math}
as above, and 

(b)
\begin{math}
  \frac{1}{\Lambda} \sum_{f \in \mathcal{F} } p(\vec{f})
  f(\vec{\s})f(\vec{\t}) = \delta(\vec{\s}; \vec{\t}) \,
\end{math}. 

\noindent 
Here, $\Lambda$ indicates the size of the family of functions, and we used the
obvious notation that makes the functions assume values $\pm 1$ if expressed
in terms of spins.  

The second assumption in the computation can be seen as a mean-field-like
hypothesis of independence of spins belonging to different clauses. It can be
argued as follows. Differently from the evaluation of
$\overline{\mathcal{N}}$, the computation of $\overline{[\mathcal{N}]^2}$
depends both on the class of functions and on the underlying network. The
essential problem is that while the function nodes are independent random
variables, the variable nodes clustered by functions can be repeated. Thus,
one has to answer to the question: how many distinct variables $n_{v}$ are
connected to $r$ constraints? For small r, we can estimate that $n_{v} \simeq
r \cdot k$, while for $ r \sim M, \ n_{v} \simeq \frac{M}{\gamma}$. These
extremes set the ``fork'' of values for which our estimate is consistent and
robust.
In these conditions, we obtain the scaling as $4^{\Delta_{C}}$, in the
relevant regime of the phase diagram $ 1/K < \gamma < \gamma^{*}$, with
$\gamma^{*} > \gamma_{c}$. This enforces the self-averaging property.
The same procedure can be carried with the k-Sat model yielding no
self-averaging for the number of satisfying states.





\section{Different Connectivities}
\label{sec:multipoiss}

\begin{figure}[htbp]
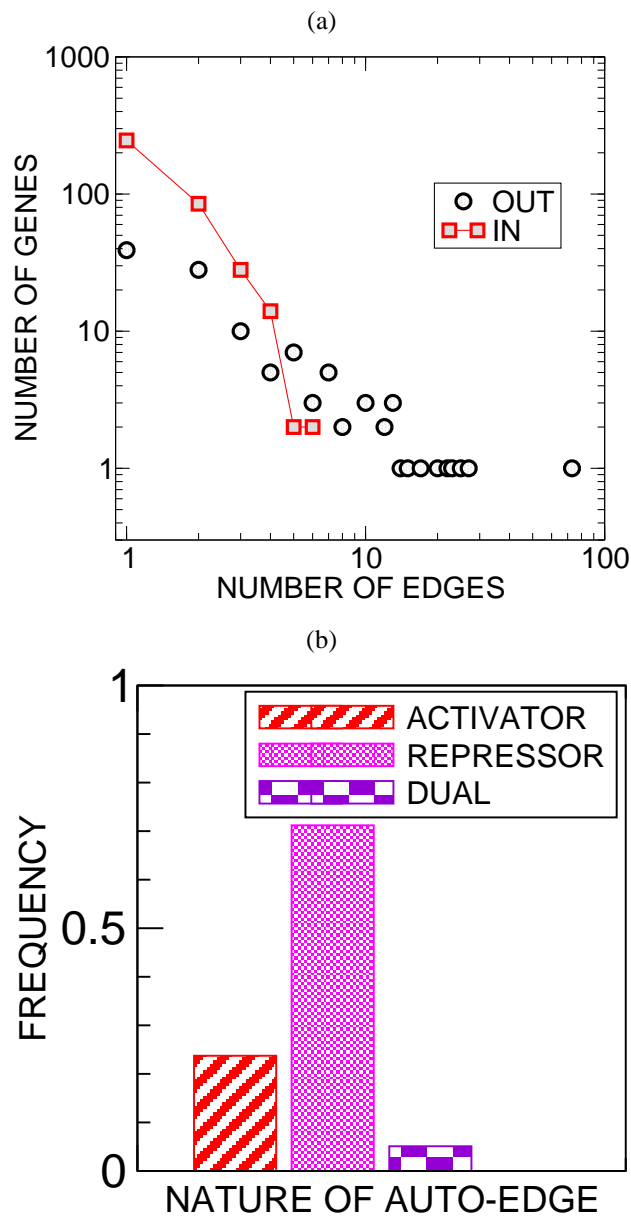

  \centering
  \subfigure[]{\includegraphics[width=.5\textwidth]{INOUTedges}}
  \subfigure[]{\includegraphics[width=.5\textwidth]{ar}}
  \caption{Data inferred from the E.~coli transcription network. (a) Degree
    distribution. (b) Activity of autoregulators. See also~\cite{MC03}}.
   \label{fig:colidati}
\end{figure}

So far we have discussed the idealized case where the in-ward connectivity,
the number of transcription factors controlling one gene, is fixed. In that
case, only the out-ward connectivity, or the number of genes controlled by a
transcription factor, can fluctuate.
A biologically more realistic case is when both the inward connectivity $K$
and the outward connectivity $c$ vary along the network, and the decay of the
latter is slower (see Fig.~\ref{fig:colidati}a). 

Considering $p(k|c) = \frac{(k\gamma)^c}{c!} e^{-(k\gamma) }$, the conditioned
probability that a variable is in $c$ clauses of the $k$ kind, we have $\pi(c)
= \sum_k\frac{(k\gamma)^c}{c!} e^{-(k\gamma) }\cdot p(k)$. The leaf removal
algorithm can be applied separately to sets of clauses with a given
connectivity, defining $ N_C \equiv <N_C>_{k}$ and $M_C \equiv <M_C>_{k}$,
where $<X>_{k}= \sum_p p(k) \cdot X(k)$.
Choosing $p(k) = Z^{-1}(\nu) e^{-\nu k}$ does not affect the exponential
asymptotic decay of $\pi(c)$ for large $c$. 

\begin{figure}[htbp]
  \centering
  \includegraphics[width=.75\textwidth]{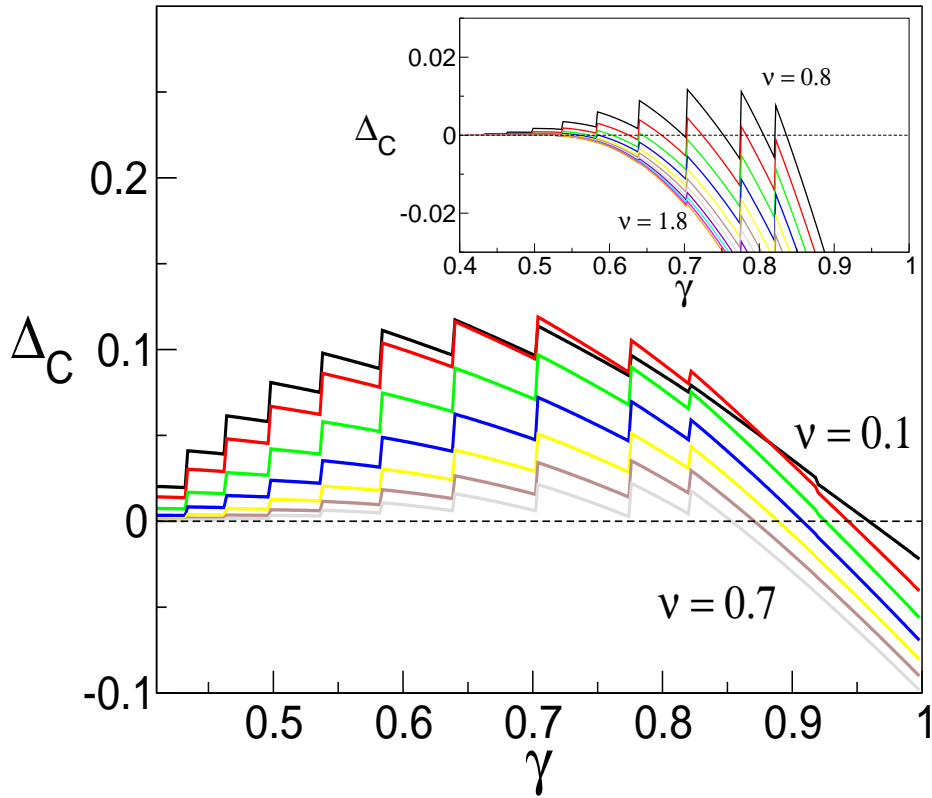}
  \caption{$\Delta_C$ as a function of $\gamma$ for different values of $\nu$
    in the multi-Poisson case. The discrete jumps are due to the onset of HARD
    phases for the different values of $k$. $\Delta_C$ can become negative
    many times, giving rise to reentrant UNSAT phases. The figure refers to a
    connectivity distribution with a cutoff at $k = 12$} 
  \label{fig:mp-delta}
\end{figure}
\begin{figure}[htbp]
  \centering
   \includegraphics[width=.75\textwidth]{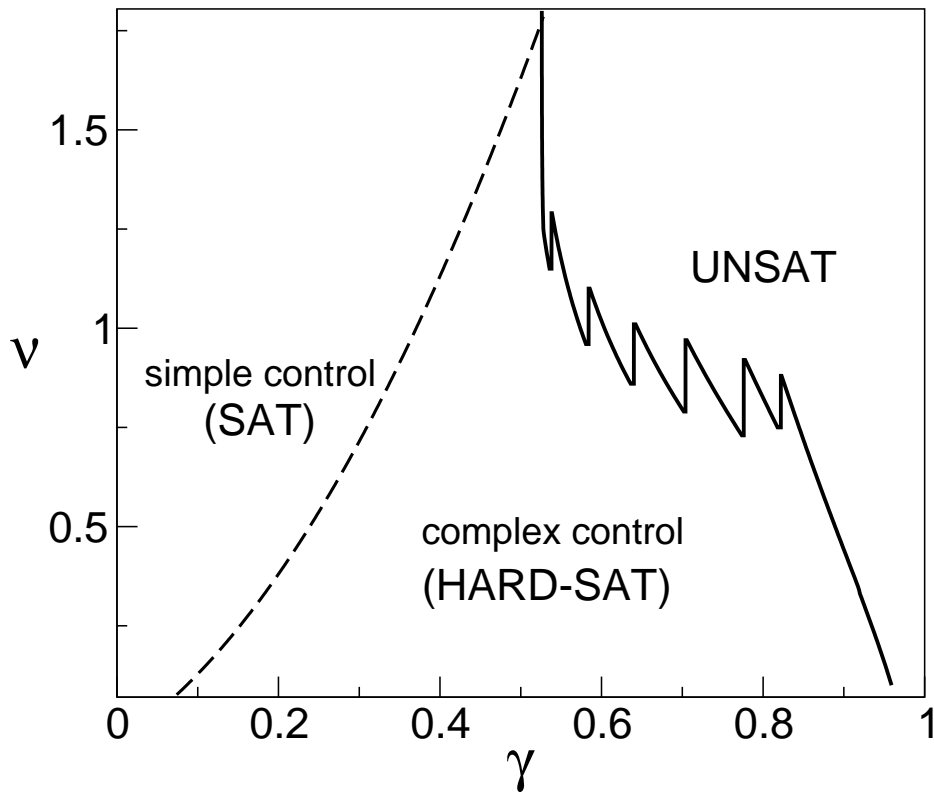}
   \caption{Phase diagram $\gamma - \nu$ for the multi-Poisson case. The dashed
     line represents the mean value of the numerically evaluated critical
     parameter $\gamma_d (\nu)$ for the SAT-HARD transition of network
     realizations with $N=3 \times 10^4$.}
  \label{fig:mp-pd}
\end{figure}

To show this, let us construct the graph with the mentioned properties 
\begin{enumerate}
\item The probability to have a clause with $k$ elements is $p(k) =
  Z^{-1}(\nu) e^{-\nu k}\quad (k>1) $.
\item $p(k|c) = \frac{(k\gamma)^c}{c!} e^{-(k\gamma) }$ is the conditioned
  probability that a variable is in $c$ clauses of the $k$ kind.
\item $\pi(c) = \sum_k\frac{(k\gamma)^c}{c!} e^{-(k\gamma) }\cdot p(k)$.
\end{enumerate}
and compute $\pi(c)$. 
Setting $\xi = \gamma + \nu$,
\begin{displaymath}
  \pi(c) = \frac{\gamma^{c}}{c!} \left( -e^{-\xi} +  \mathcal{Z}[k^{c}]
  \right) \ \ ,
\end{displaymath}
where $\mathcal{Z}[f(c)] = \sum \frac{f(c)}{z^{c}}$ is the Z-transform, and
for us $z = e^{\xi}$. Now, $ \mathcal{Z}[p^{c}] = \textrm{Li}_{-c}(z^{-1})$,
where Li is the polylogarithm, which can be defined for negative integers as 
\begin{displaymath}
  \textrm{Li}_{-c}(z) := \frac{1}{(1-z)^{c+1}} \sum \eulang{c}{i} r^{c-i} \ \ ,
\end{displaymath}
where  $ \eulang{c}{i} $ are Euler's numbers.
This gives a condensed expression for $\pi(c)$:
\begin{displaymath}
  \pi(c) = \frac{\gamma^{c}}{Z c!} \left\{  \textrm{Li}_{-c}(e^{-\xi}) -
    e^{-\xi} \right\} \ \ . 
\end{displaymath}
It can be easily checked that, for large $c$, this function decays
exponentially, after having reached a maximum.

We can call this case multi-Poisson, as the graph is a superimposition of
graphs that follows a Poisson distribution, each graph having in turn a fixed
clause-connectivity and a Poisson variable-connectivity. 
%
%
The behavior of GR1 on such a topology is   different from the
fixed connectivity case.  The main reason for this is that, while
$\Delta_{C}(\gamma)$ is still locally decreasing, many new discontinuities
emerge, due to the influence of clauses with different connectivities. This
gives rise to two phenomena. Firstly, $\Delta_{C}$ can increase globally with
increasing $\gamma$. Indeed, it does increase after $\gamma_{d}$, to decrease
again before $\gamma_{c}$ (Fig.~\ref{fig:mp-delta}). After the onset of the
complex control phase, the complexity initially increases (step-wise), reaches
a maximum, and then decreases monotonically.  This has an influence on the
number of observed states as a function of $\gamma$.  Secondly, $\Delta_{C}$
can become negative and then jump back to a positive state, creating a
reentrant HARD-SAT phase (Fig.~\ref{fig:mp-pd}).
We are currently studying ways to extend our calculation of the mean number of
compatible states and the width of its distribution on graphs with more
general connectivities.

\section{An Example from an Experimental Setting}
\label{sec:concr}

The results described so far focused on the typical behavior of GR1 as a
formal model for a genetic network. To resume them, we can predict the
existence of a core of variables, the CCC, which determines the behavior of
the system. The phase diagram of the system contains two regimes of gene
control, simple and complex.  In the complex control phase, the free genes of
the core control $O(N)$ other genes. These phases also depend on connectivity.
On the other hand, a very important question is how to relate them to concrete
systems.  There are many possibilities in this direction that we are currently
exploring. In this section, we will discuss a first attempt. Specifically, we
will make use of the data set for the structure of the E.~coli
transcription network from the RegulonDB database~\cite{SSG+01}, with the
modifications of~\cite{SMM+02}. The goal is to apply the leaf removal
algorithm using the information contained in the data set.

The data set consists of an annotated graph, where the signal integration
functions are described as sets of annotated links. The annotations consist in
three modes of activity: activation, repression, and ``dual'' activity
(meaning that the activity depends on the context). The data on the
combinatorial activity of transcription factors are not part of the set. For
this reason, in what follows we will ignore the annotations, concentrating on
the study of random GR1 realizations on the given experimental network
structures.
Considering the connectivity matrix $C_{ij}$ defined as
\begin{displaymath}
  C_{ij} = \left\{
  \begin{array}{lll}
    1 ; \ \  \textrm{Gene j regulates gene i} \\
    0 ; \ \  \textrm{Otherwise}
  \end{array}
  \right. \ ,
\end{displaymath}
a ``leaf'' corresponds to a column containing only zeros.  An iteration of the
algorithm removes these columns, together with the corresponding lines. Note
that the leaf removal algorithm is not guaranteed to preserve the network
structure as in the abstract cases discussed above.

During the iterative leaf removal procedure, one is confronted with an
important choice, concerning how to deal with autoregulators. These create a
problem, as, for particular assignments of the functions they create trivial
contradictions.
In reality, this self-contradiction is inexistent, as negative
autoregulations are known to play the role of controlling the overexpression
of a particular gene.  A standard, and the simplest, way to avoid the problem
is simply to eliminate the autoregulations, and impose that the diagonal of
$C_{ij}$ is zero.
However, this total cancellation is not biologically motivated - as
autoregulations might reflect some global properties of the system, other than
control of overexpression. To clarify, let us consider a gene that regulates
itself and is regulated by some others (``rest''), that is a ``regulated
autoregulator'' (RAR). It is then subject to the constraint $\s_0 = f ( \s_0 |
\textrm{rest} ) = A (\textrm{rest})\s_{0} + B (\textrm{rest})$.  If
$A(\textrm{rest}) = 0$ the gene is regulated simply, for $B\in \big\{-1,1
\big\}$ and the autoregulation is irrelevant. Conversely, when $B
(\textrm{rest}) = 0$, the autoregulation plays a role, but if $A
(\textrm{rest} ) = -1$ the system is UNSAT.

To solve this problem, we propose a way to keep the role of autoregulators
into account, while at the same time avoiding the trivial self-contradiction.
In order to do this, we introduce the constraint $A (\textrm{rest} ) = 1$ that
codes for the avoidance of trivial contradictions. With this technique, we aim
to save the autoregulation role, while taking into account the notorious fact
that auto-inhibitions cannot be represented with Boolean variables. We can
call this the ``RAR hypothesis''.  We will see that this hypothesis brings to
a different final result. The same reasoning can be carried for GRq-type
variables.
Assuming the RAR hypothesis, the problem becomes a mixed optimization
problem, that includes the usual GR1 constraints, plus a set of Sat-like
constraints that come from the $A_{n} (\textrm{rest} ) = 1$ conditions on the
RARs. 

\subsection{The E.~coli Transcription Network}

\begin{figure}[htbp]
  \centering
  \includegraphics[width=.8\textwidth]{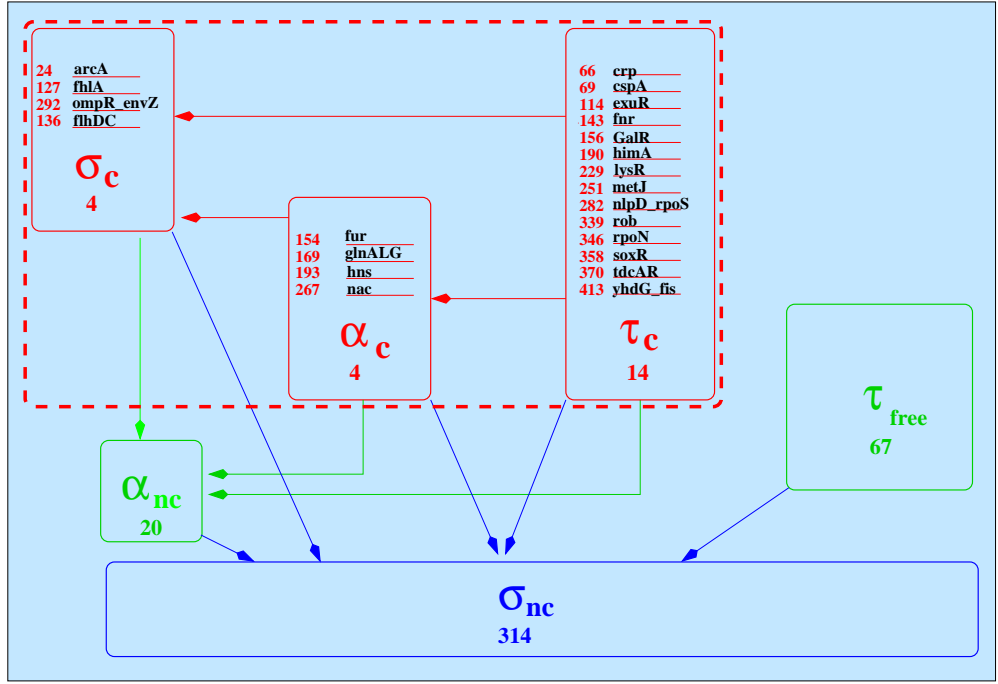}
  \caption{The CCC of E.~coli with the RAR hypothesis. It contains 22
    variables (of which 14 are free or regulated only by themselves, 4 are
    non-free, and 4 are RARs) and 22 constraints (of which 18 are RAR
    constraints). }
  \label{fig:colirar0}
\end{figure}

In the E~.coli data set there are 423 genes, and 59 autoregulations. Among
these, 24 are RARs (Fig.~\ref{fig:colidati}). Applying the leaf-removal
algorithm with cancellation of autoregulations leads to an empty core. This
means that the system finds itself in the simple control, SAT phase.  However,
the application of the RAR hypothesis leads to a non-empty core
(Fig.~\ref{fig:colirar0}).

The genes in the core can be divided in three different classes, free, which
we will denote by $\tau$, non-free ($\sigma$), and RARs, or ($\alpha$). The
core contains a total of 22 variables (of which 14 are free or regulated only
by themselves, 4 are non-free, and 4 are RARs) and 22 constraints (of which 18
are RAR constraints). 
Biologically speaking, these core gene include some ``global regulators'', or
transcription factors with a high out-ward connectivity~\cite{MC03}, including
(a) the sigma factors rpoS and rpoN, (b) proteins belonging to the family of
the DNA bending global regulator crp (c) himA, or IHF, another DNA bending
factor.
More interestingly, also lower connectivity proteins, connected to metabolism
(e.g. respiratory control and iron transport), and to structural tasks (e.g
synthesis of the flagellum) are found in the core.

\begin{figure}[htbp]
  \centering
  \subfigure[]{\includegraphics[width=.75\textwidth]{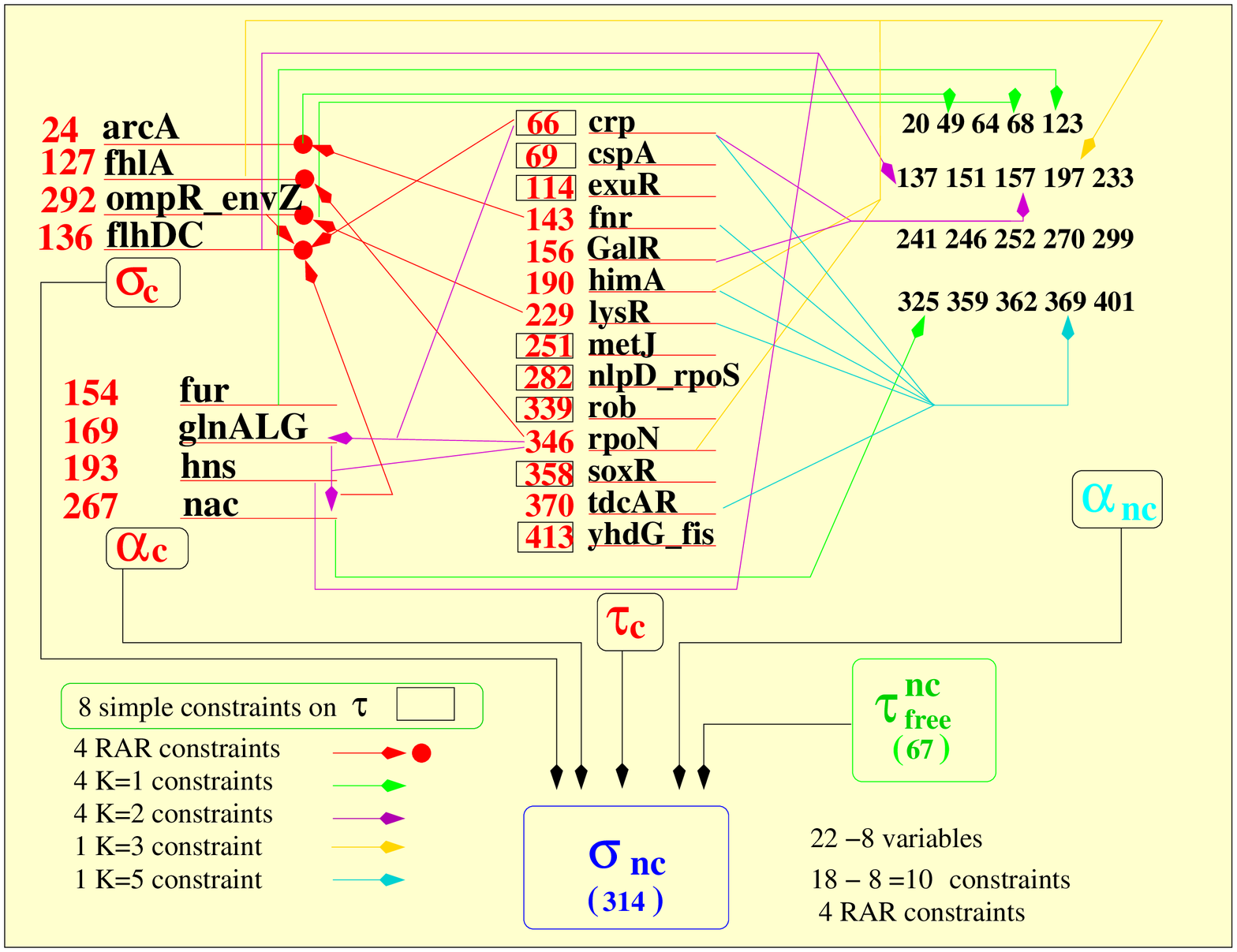}}
  \subfigure[]{\includegraphics[width=.75\textwidth]{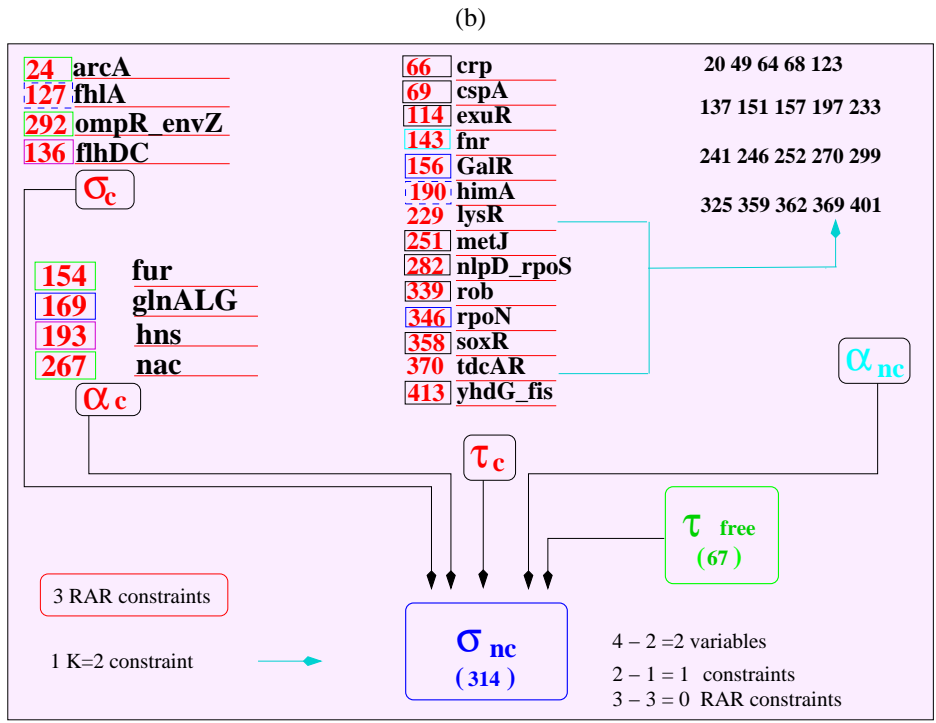}}
  \caption{Solution of the general optimization problem on the core variables
    of E.~coli in the RAR hypothesis. In this procedure, variables are fixed
    with respect to each other according to the constraints that connect them.
    (a) Second step of the computation (b) Last step of the computation.}
  \label{fig:colirar12}
\end{figure}

The residual optimization problem on the core variables is small and simple
enough to be solved for general functions, as exemplified in
Fig.~\ref{fig:colirar12}. The final solution gives only two states, after
having fixed the free genes.

What is the meaning of the CCC in the RAR hypothesis, if any? The answer can
come from two directions: simulations and experiments.  For the numerical
case, one must study how fixing the core variables affects the reach of a
fixed point or a steady state in a Boolean network.  Our simulations on both
asynchronous spin-flip and synchronous update dynamics show that, fixed some
random functions on the whole network, the core free genes control a larger
set of genes than the non-core ones (these results will be published
elsewhere~\cite{CLB04}). This is an indication that the CCC found with the RAR
hypothesis might have some significance.  The same feature can be tested with
microarray expression experiments.




\section{Conclusions}

In conclusion, we have presented and discussed a novel conceptual framework
for the equilibrium modeling of large scale transcription networks. In its
most general formulation, our approach is directly connected to the
Shea-Ackers model for the \emph{cis-} regulatory region of a gene, and
consists of a compatibility analysis of the constraints established by the
signal integration functions.

The advantage of this approach is that it allows to separate issues related to
the dynamics of the network from the basic logic structure that underlies it.
Obviously, dynamics is a very important factor of a real biochemical network,
possibly the most important.  On the other hand, we feel that the
disentanglement the two aspects might lead to further insight.
In the spirit of theoretical computer science, any dynamics superimposed on
GR1 can be seen as an algorithm.  The problem becomes then the following. How
effectively does a given algorithm, modelling chemical kinetics explore
configuration space?  Naturally, this addition may carry intricate issues,
connected to the nonequilibrium nature and the asymmetry of the interactions.
These issues are particularly complex if one wants to add a coarse-graining
of time, as is commonly done in Kauffman networks.
In absence of an explicit knowledge of the emergent time scales involved in
the dynamics, we feel ours is an appropriate approach.  Particularly in the
Boolean approximation, GR1, which we treat here.  
%

%

From a general, speculative standpoint, our model shows that the ``biological
complexity'' is not simply measured by the number of genes. For a
transcription network, a more proper indicator is $\Delta_{C}$. Interestingly,
for GR1 this coincides exactly with what is called the ``computational''
complexity, $\Sigma$.
Looking at the phase digram, $\Sigma$ depends on the order parameter
$\gamma$, or - loosely - on the number of transcription factors per gene.  At
fixed number of genes, it is known that this quantity increases in bacteria
that need to react to more environments~\cite{CdL03}.
Imagining that prokaryotes, being unicellular, naturally find themselves in a
simple control phase, our phase diagram predicts an intrinsic limit to this
adaptation process, represented by the phase boundary with the HARD-SAT.
Considering varying $N$, one may wonder why, in real organisms, a small
$\gamma$ is correlated with a small $N$~\cite{vN03}. A possible answer to this
question is the following. With large $N$ and small $\gamma$, the system is
shifted to the SAT phase, and therefore needs to explore a very big
configuration space without sufficient ``guidelines''. In other words, the
available configurations are too many to be reached in reasonable time by the
dynamics.

Similar considerations can be carried for the \emph{width} of the distribution
of satisfying solutions. The fact that the self-averaging property holds
indicate that this is negligible in the thermodynamic limit.  On the other
hand, the typical value of $N$ for a living system is in the range
$10^{3}-10^{5}$.  While being large for detailed modeling, this is a
smaller number than the size of the typical system treated with statistical
mechanics. Thus, the effects of the system size are expected to be important.

Considering the phase diagram in Fig.~\ref{fig:pd}, the complex control phase,
having general residual dynamics, matches a qualitative feature of many cells,
where some genes are constantly expressed, and the rest vary.
On the other hand, the dynamical slowing down characteristic of any glassy
phase raises an issue that must be solved by the chemical dynamics of the
cell. In analogy with Kauffman's ideas, the breakdown in many different
attraction basins might be interpreted as epigenesis.  That is, in the
HARD-SAT phase there will be typically many cell types. How many, is
determined by the complexity $\Sigma$, which is directly measured by our
$\Delta_{C}$.  While for fixed $K$ this quantity simply decreases with
increasing $\gamma$, its behavior is more interesting in the multi-Poisson
case.
However, this remains an open issue which has to be regarded with more detail.
The experimental scaling of the number of cell types is sub-linear in the
number of genes $N$~\cite{Kau93,Kau04}.  In the fixed $K$ case, GR1 gives
exponential scaling with $N$ at fixed $\gamma$ in the complex control phase.
On the other hand, the results of random-GR1 are perhaps more easily related
to the number of species times the number of cell types at equal number of
genes. In the same way, the behavior of $\overline{[\mathcal{N}]^2}$ with $N$
should roughly predict the scaling of the \emph{variability} in the total
number of cell types for all the species with equal number of genes $N$.
According to GR1, this quantity should vanish in the thermodynamic limit.

To be biologically useful, the model has to deal with the details of an
individual realization the system.  In this respect, an advantage of the leaf
removal algorithm is that it transforms a problem related the states of
variables on a graph, the gene expression patterns, into a problem regarding
the \emph{structure} of the graph.  This is particularly of interest as long
as the data regarding the activity of function nodes are only partially known.
For example, the first application to the E.~coli core, in the RAR hypothesis
leads to interesting results, that have a numerical counterpart and might be
tested with expression correlation data. The application to more, larger, data
sets and to other forms of regulation might lead to further insight.
Notably, some of the core variables do not have a high connectivity.  This is
an indication that additional, global, properties of the network structure
other than local order parameters must contribute to establish the hierarchy
of states in configuration space. 

Finally, besides the extensions to the work presented here, we believe the
framework of GR might be used as a setting for many different problems
involving fairly large networks, from evolutionary models to regulation
network optimization, from network inference to design. It will be potentially
useful in the years to come, as more and more data will be available from
high-throughput experiments.


%

%

%
%




%

\begin{acknowledgments}
  We would like to acknowledge interesting discussions with L.~Finzi,
  A.~Sportiello, J.~Berg, M.~Leone, M.~Caselle, P.R.~ten~Wolde, R.~Zecchina.
\end{acknowledgments}

\begin{appendix}

\section{Self-averaging property of GR1.}
\label{sec:self-aver-prop}

The following paragraphs describe the calculation of the width of the
distribution of $\mathcal{N}$. By definition,
$$\overline{[\mathcal{N}]^2}=\sum_{\vec{\s},\vec{\t}}\sum_{C}p(C)\sum_{f^1 \in
  \mathcal{F} }p(f^1)\sum_{f^1 \in \mathcal{F}}p(f^2)...\sum_{f^M\in
  \mathcal{F}}p(f^M) \cdot$$
$$\cdot \prod_{m=1}^M \d
\big(1-f^m(\s_{n(1,m)},\cdots,\s_{n(k,m)})\cdot\s_{n(0,m)}\big) \d
\big(1-f^m(\t_{n(1,m)},\cdots,\t_{n(k,m)})\cdot\t_{n(0,m)}\big) \ \  ;$$

thus, 
$$\overline{[\mathcal{N}]^2}= \sum_{\vec{\s},\vec{\t}}\sum_{C}p(C)
\prod_{m=1}^M \Big(\sum_{f^m  \in \mathcal{F}}p(f^m)
 \d \big(1-f^m(\s)\cdot\s_{n(0,m)}\big)
 \d \big(1-f^m(\t)\cdot\t_{n(0,m)}\big)
\Big) \ \ .$$

For fixed states $\vec{\s}$ and $\vec{\tau}$, we can write the factors of the
product above as
$$\Big(
 \d \big(1-\s_{n(0,m)}\big)
 \d \big(1-\t_{n(0,m)}\big)\cdot A(\s_m ;\tau_m)+ 
\d \big(1-\s_{n(0,m)}\big)
 \d \big(1+\t_{n(0,m)}\big)\cdot B(\s_m ;\tau_m) + $$
$$\d \big(1+\s_{n(0,m)}\big)
 \d \big(1-\t_{n(0,m)}\big)\cdot C(\s_m ;\tau_m)  +
\d \big(1+\s_{n(0,m)}\big)
 \d \big(1+\t_{n(0,m)}\big)\cdot  D(\s_m ;\tau_m) 
\Big) \ \ ,$$

Where $A$, $B$, $C$, $D$, are the weights of the functions $ f \in
\mathcal{F}$ such that, respectively
\begin{displaymath}
\left \{ \begin{array}{lllll}
\textrm{$ A(\s_m ;\tau_m) $}&\textrm{$\leftarrow$} & \textrm{$f(\s \in m)=1$}
&\& & \textrm{$f(\t \in m)=1$} \\ 
\textrm{$ B(\s_m ;\tau_m)$}&\textrm{$\leftarrow$} & \textrm{$f(\s \in m)=1$}
&\& & \textrm{$f(\t \in m)=-1$} \\  
\textrm{$ C(\s_m ;\tau_m)$}&\textrm{$\leftarrow$}  & \textrm{$f(\s \in m)=-1$}
&\& & \textrm{$f(\t \in m)=1$} \\ 
 \textrm{$ D(\s_m ;\tau_m)$}&\textrm{$\leftarrow$} & \textrm{$f(\s \in m)=-1$}
 &\& & \textrm{$f(\t \in m)=-1$} \\ 
\end{array}\right.
\end{displaymath}

Now, as $A+B+C+D =1$, and $A+B = A+C = \rho$, we can write,
choosing $\rho=1/2$ 

$$\sum_{C } \sum_{\vec{\sigma} \vec{\tau}} \prod_{m} $$
$$\left\{\right. A(\s_{m},\t_{m}) \cdot \big[\d \big(1-\s_{n(0,m)}\big)\d
\big(1-\t_{n(0,m)}\big) - \d \big(1+\s_{n(0,m)}\big)\d
\big(1+\t_{n(0,m)}\big)] \ \ +  \ \ \ \ (\mathcal{A})$$
$$ + \frac{1}{2} \big[ \d \big(1-\s_{n(0,m)}\big)\big(1+\t_{n(0,m)}\big) + \d
\big(1+\s_{n(0,m)}\big)\big(1-\t_{n(0,m)}\big) \big] \left. \right\} \ \ \ \
(\mathcal{R})$$

The product on $m$ gives rise to $2^{M}$ terms of the kind
$$
\prod_{k=1}^{r} \mathcal{A}^{(k)} \prod_{k'=r+1}^{M} \mathcal{R}^{(k')} \ \ 
,$$
where $\mathcal{A}$ and $\mathcal{R}$ indicate factors of the two types in
the previous expression. For every $r$ there are $\su{M}{r}$ terms of this
kind in the sum. Applying the properties that characterize our family of
functions, we find $A= \sum_{f \in \mathcal{F}}\big[
\big(\frac{1+f(\s)}{2}\big) \big(\frac{1+f(\t)}{2}\big)\big]=
\frac{1}{4}\big[1 + \d(\s,\t)\big] \ \, $ 

With this consderation, the sum over the configurations $\vec{\sigma}
\vec{\tau}$ can be simplified. It involves the product
$$ \prod_{m} $$
$$\left\{ \right. A(\s,\t) \cdot \big[\s_{n(0,m)} \t_{n(0,m)}\big] \ \ + \ \ \ 
  \ \ \ \ \ (\mathcal{A})$$
$$ + \frac{1}{4} \big[1 - \s_{n(0,m)} \t_{n(0,m)} \big] \left.
\right\} \ \ \ \ \ \ \ \ \ \ \ \  (\mathcal{R})$$

Let us now distinguish again between free genes and non-free ones, which are
outputs of some signal integration function.  The sum over non-free genes is
such that (i) there is a contribution $ \mezzo^{r}$ due to the Kronecker
deltas in the $\mathcal{A}$ part and (ii) if a type $\mathcal{R}$ non-free
$\s_{n(0,k')}$ or $\tau_{n(0,k')}$ variable appears in the input of a type
$\mathcal{A}$ clause the contribution is zero.  A little thought leads to the
conclusion that the non-free genes sum up to the term
\begin{displaymath}
  \left(\mezzo\right)^{r} \left(1-\gamma + \frac{r}{N}\right)^{kr}
\end{displaymath}

The sum over the $2(N-M)$ free genes would give a $4^{\Delta}$ contribution in
case of complete independence.  However, a delta function on the input genes
reduces the double sum to a single one. To estimate this contribution one has
to evaluate the probability that two free genes appear as input of a type
$\mathcal{A}$ clause. In a mean-field like estimate, this is $ kr
\frac{N-M}{N-M +r}$, leading to the contribution
\begin{displaymath}
  4^{\Delta} \cdot 2^{-\frac{kr}{1+\frac{r}{\Delta}}}
\end{displaymath}

The $4^{\Delta}$ term factors out of everything, and alone would give the
desired self-averaging property. It remains to evaluate the sum over $r$.
Restricting the sum over the core genes, and evaluating it with a saddle point
method leads to the minimization of the free-energy-like functional
\begin{displaymath}
  G(x) = x \log x + (1-x) \log (1-x) + \log(2) x\left( 1 + 
    \frac{k(1-\gamma)}{1-\gamma(x-1)}\right) - k x
    \log\left(1-\gamma(x-1)\right)  \ ,
\end{displaymath} 
where $x \in [0,1]$, and $\gamma := \frac{M_{C}}{N_{C}}$.
Minimization of this functional always leads to the solution $x=0$, with the
exception of the regions: $\gamma < 1/K$ and $\gamma > \gamma^{*}$, where
$\gamma^{*}$ is a threshold that always lies in the UNSAT region. For $k=3$,
$\gamma^{*} \simeq 0.9722$.

\end{appendix}

\end{document}